\documentclass[twocolumn,twoside,slac_two]{revtex4}
\usepackage{graphicx}
\usepackage{fancyhdr}
\usepackage{graphicx}
\usepackage{amssymb,amsmath}

\usepackage{enumerate}
\usepackage{subfigure}
\usepackage{rotating}

\newcommand{\met}    {\mbox{$\protect \raisebox{.3ex}{$\not$}E_{T}$}}

\makeatletter

\newcommand{\Rmnum}[1]{\expandafter\@slowromancap\romannumeral #1@}
\makeatother

\begin{document}


\title{Prospects for measuring \textbf{$t\bar{t}$} production cross-section at $\sqrt{s}=10$ TeV using the likelihood method with the ATLAS detector}
%

\author{Dilip Jana}
\affiliation{University of Oklahoma, Norman, OK 73019, USA}

\begin{abstract}
Due to the large  $t\bar{t}$  production cross-section at the LHC energies, the ATLAS experiment is expected to have enough statistics to measure $t\bar{t}$ cross-section even at initial luminosities. Recent studies performed in ATLAS on the development of $t\bar{t}$  cross-section measurements in the lepton+jets channel at $\sqrt{s}$=10~TeV using the likelihood method will be discussed. The expected statistical and systematic uncertainties for the cross-section measurement using the likelihood method are evaluated for an integrated luminosity of 50 pb$^{-1}$ of Monte Carlo (MC) simulated data. Measurements with data that will be collected in the first year of the LHC operation are emphasized.

\end{abstract}

\maketitle

\thispagestyle{fancy}


\section{Introduction}
The top quark was discovered in 1995 at Fermilab in the pair production mode ($t\bar{t}$ events) through strong interactions [1]. At the LHC, $t\bar{t}$ production takes place at small momentum fractions of the proton, $x \approx 10^{-2}$ where the gluon parton distribution function (PDF) is large. Therefore, the gluon scattering process ($\approx 90 \%$ at 10 TeV) dominates over quark-antiquark scattering in $t\bar{t}$ production at the LHC. Because of much larger parton luminosities (dominated by the gluon PDF), the $t\bar{t}$ production cross-section, $\sigma_{t\bar{t} }$, at the LHC is significantly larger than at the Tevatron. Assuming $m_{t}=172.5$ GeV and using CTEQ6.6 and CTEQ6.5 PDFs [2], the $t\bar{t}$ cross-section at 10 TeV is 400 pb$\pm$11$\%$ at next-to-leading order (NLO) and 400 pb$\pm$6$\%$ at next-to-next-to-leading order (NNLO). Any deviation in the measured cross-section from the theoretical value can be considered an indication of the presence of physics beyond the Standard Model. Measurements of the $t\bar{t}$ cross-section in different channels ($e+$jets, $\mu+$jets, $\tau+$jets, di-leptons) provide a more stringent test of Standard Model predictions, since new physics may modify relative fractions of these channels with respect to the total cross-section. Top quark events will be a major background for Higgs and supersymmetry (SUSY) searches. Therefore, accurate knowledge of the $t\bar{t}$ cross-section is crucial during the early operation of the LHC to discover new physics.

In the Standard Model, the top quark decays almost exclusively through the $t\rightarrow Wb$ decay. From the experimental point of view, we consider the case when one of the $W$'s decays leptonically, and the other decays hadronically ($\mathrm{t\overline{t}\rightarrow W W b \overline{b} \rightarrow l \nu j_1 j_2 b \overline{b}}$ with $l=e,\mu$ and $j_{1}, j_{2}$ are light jets). This is known as a $semi$-$leptonic$ $t\bar{t}$ decay. This allows us to trigger events efficiently using the single lepton trigger. The final state of the semi-leptonic channel consists of a high-$p_T$ charged lepton, missing energy and four jets. This decay channel is usually called lepton+jets, and we will use this terminology in this paper.

Several methods for the $t\bar{t}$ inclusive production cross-section measurement have been developed and carefully studied by the ATLAS collaboration [3]. All of these methods have the goal of an early $t\bar{t}$ cross-section measurement. Measuring the $t\bar{t}$ cross-section using the likelihood method can be complementary to other  analyses and therefore provides a valuable cross check of the early results. This method uses shapes of different kinematic variable distributions in $t\bar{t}$ and background events. We choose kinematic variables which have good separation power between $t\bar{t}$ and backgrounds. We optimize those discriminating variables to reduce statistical and systematic uncertainties. We have not used b-tagging [3] in this analysis.


\section{The Likelihood Discriminant}
The discriminating function can be generally written as follows:
\begin{equation}
L = \frac{S(x_1,x_2,...,x_N)}{S(x_1,x_2,...,x_N)+B(x_1,x_2,...,x_N)} 
\end{equation}
where $x_N$ are discriminating variables and $S$ and $B$ are N-dimensional probability density functions for signal and background, respectively. We can assume to good approximation that the N discriminant variables are uncorrelated. Therefore, each multidimensional probability density function can be expressed as a product of one dimensional probability density functions:
\begin{equation}
\begin{split}
L &= \frac{\Pi_{i=1}^{N} S_i(x_i)}{\Pi_{i=1}^{N} S_i(x_i) + \Pi_{i=1}^{N} B_i(x_i)} \\
&=\frac{\exp(\sum_{i=1}^{N}\ln\frac{S_i(x_i)}{B_i(x_i)})}{\exp(\sum_{i=1}^{N}\ln\frac{S_i(x_i)}{B_i(x_i)})+ 1}\\
\end{split}
\end{equation}

Denoting the fit to the logarithm of $S$ and $B$ as $(\ln\frac{S}{B})_{i}^{fitted}(x_{i})$, we can write the likelihood discriminant as
\begin{equation}
L= \frac{\exp(\sum_{i=1}^{N}(\ln\frac{S}{B})_{i}^{fitted}(x_{i}))}{\exp(\sum_{i=1}^{N}(\ln\frac{S}{B})_{i}^{fitted}(x_{i}))+ 1}
\end{equation}


\section{Cross-section extraction procedure}
The $t\bar{t}$ production cross-section is given by
\begin{equation}
\sigma_{t\bar{t}}= \frac{N_{t\bar{t}}}{Br\cdot L\cdot \epsilon_{sel}}
\end{equation}
where $N_{t\bar{t}}$ is the number of selected $t\bar{t}$ events, $Br$ is the branching ratio for the final state considered, $L$ is the integrated luminosity and $\epsilon_{sel}$ is the selection efficiency.

\section{Reconstruction of physics objects}
We have used semi-leptonic decays of $t\bar{t}$ events ($\mathrm{t\overline{t}\rightarrow l \nu j_1 j_2 b \overline{b}}$ with $l=e,\mu$) for this analysis.
In this section, we define basic reconstructed objects (electrons, muons, jets and missing transverse energy) used in this analysis. These definitions are based upon the standard reconstruction algorithms recommended in ATLAS. The pseudorapidity is defined as $\eta$ = -$\ln$ tan$(\frac{\theta}{2})$ where $\theta$ is the polar angle from the beam axis. The azimuthal angle $\phi$ is measured around the beam axis. We define separation of two objects in $\eta - \phi$ space as $\Delta R = \sqrt{(\Delta \phi^{2} + \Delta \eta^{2})}$.


\subsection*{Electrons}
Electrons are reconstructed in the inner tracker and in the calorimeter of the ATLAS detector. Electrons are required to have pseudo-rapidity in the range  $|\eta| \leq$ 2.47 and $p_{T} > 20$ GeV. Furthermore, if an electron is found in the calorimeter crack region $1.37 < |\eta| <1.52$, the electron is discarded. The electron has to be isolated based on the calorimeter energy: the additional transverse energy ($E_{T}$) in a cone with radius $\Delta R = 0.2$ around the electron axis is required to be less than $6$ GeV.

\subsection*{Muons}
Muons are reconstructed by combining information from the inner detector and the muon spectrometer. The muons used in this analysis must have minimum transverse momentum $p_{T} > 20$ GeV and are required to lie in the pseudo-rapidity range $|\eta| < 2.5$. Muons are isolated based on the calorimeter energy: the additional transverse energy $E_{T}$ in a cone with radius $\Delta R = 0.2$ around the muon is required to be less than $6$ GeV. In order to remove muons coming from decays of hadrons inside jets (i.e., $B$-hadrons originating from the $b$-quark in top quark decays), muons which are within a cone of size  $\Delta R < 0.3$ from a jet are removed.


\subsection*{Jets}
Jets are reconstructed using the standard ATLAS cone  algorithm [3] in $\eta - \phi$ space, with a cone radius of 0.4. The jets used in this analysis are required to have minimum transverse momentum $p_{T} > 20$ GeV and are required to lie in the pseudo-rapidity range $|\eta| < 2.5$.  Jets which overlap with electrons within a cone of size $\Delta R < 0.2$ are discarded.




\subsection*{Missing Transverse Energy (\textbf{$\met$})}

For the calculation of missing transverse energy $\met$, we determine the sum of the following components: the contribution of cells in identified electron or photon clusters, the contribution of cells inside jets, the contribution of cells in topological clusters outside identified objects, the contribution from muons and the cryostat correction [3]. The sum of total transverse energy in semi-leptonic top quark events is about 500 GeV, which gives a typical $\met$ resolution of the order of 10 GeV [3].

\section{Selection of lepton + jets events}

We identify semi-leptonic $t\bar{t}$ events by requiring that either the single isolated electron or single isolated muon trigger, both with 15 GeV thresholds, have fired. The trigger efficiencies can be measured either from MC simulation or from data ($Z\rightarrow ee$ or $Z\rightarrow \mu\mu$) [3].


Apart from the trigger requirement, the following additional event selection cuts have been applied for event selection:

\begin{itemize}
\item Exactly one isolated lepton (electron or muon) with $p_T>20$ GeV.
\item $\met>$20 GeV.
\item at least 3 jets with $p_T>$40 GeV.
\item at least 4 jets with $p_T>$20 GeV.

\end{itemize}

The fraction of all $t\bar{t}$ events passing the individual selection requirements and overall commulative efficiency are shown in Table ~I.

\begin{table}[htp]
\caption{\textbf{Cut flow efficiency for the $t\bar{t}$ decay.}}
\begin{tabular}{|c|c|c|}  
\hline
                  & $\mu$+jets (\%) & e+jets (\%) \\ \hline
Trigger        & $31.0\pm0.1$ & $25.0\pm0.1$ \\
Isolated lepton & $67.0\pm0.1$ & $70.3\pm0.2$ \\
$\met> 20 $GeV  & $91.2\pm0.1$ & $90.6\pm0.1$ \\
3 jets with $p_T>40$ GeV & $47.7\pm0.2$ & $47.5\pm0.2$ \\
4 jets with $p_T>20$ GeV & $80.0\pm0.2$ & $79.7\pm0.2$ \\ \hline
Cumulative efficiency & 7.23$\pm$0.04  &  6.01$\pm$0.04          \\ \hline

\end{tabular}
\label{tab:presel_signal}
\end{table}

\section{Backgrounds}
The lepton+jets data have a mixture of both $t\bar{t}$ events and background events ($W+$jets, multi-jet QCD, single top and di-boson). The expected numbers of $t\bar{t}$ and backgrounds in 50 pb$^{-1}$ are shown in Table~II. The major background is $W+$jets. In order to construct the likelihood template, we used discriminating variables which have good separation power between the $t\bar{t}$ and $W+$jets events. The di-boson background behaves like the $W+$jets background, and therefore it is included as part of the $W+$jets template. The  single top likelihood template behaves like the $t\bar{t}$ template. Therefore, the number of $t\bar{t}$-like events obtained with the likelihood templates fit includes the single top production. We will subtract the expected number  of single top events from the signal-like events using the theoretical cross-section of single top production. QCD backgrounds are expected to be present in the data sample due to imperfect jet and lepton identification and due to the presence of heavy flavor jets, where $b$ or $c$ quarks decay semi-leptonically. The number of expected QCD events will be determined solely from data using the matrix method [5]. The contribution of QCD events will be subtracted from the data in each bin of the likelihood distribution. 



\begin{table}[htp]
\caption{\textbf{Expected numbers of $t\bar{t}$ and major background events in 50 pb$^{-1}$ of Monte Carlo (MC) simulated data.}}

\begin{tabular}{|c|cc|cc|}
\hline 
Data sample  & $\mu$+jets & & $e+$jets & \\
             
\hline
  $t\bar{t}$         & 785 & & 653 & \\ \hline
$W(\mu\nu)+$jets & 416 & & 0 &   \\ \hline
 
$W(e\nu)+$jets   & 0    & & 297 & \\ \hline 
single top       & 56  && 52 &   \\ \hline

Di-boson          & 3 && 2 &  \\ \hline
QCD: Di-jet       & 8 && 13 &\\ \hline
\end{tabular}
\label{tab:Expect_event}
\end{table}

\section{Likelihood Discriminant variables}

For the likelihood function, we choose variables that satisfy the following
requirements:
\begin{enumerate}
\item Their distributions in data are expected to be accurately
reproduced by MC simulation;
\item They have either minimal or no dependence on the jet energy scale
corrections which could be a source of large systematic errors in early data;
\item They have either minimal or no dependence on missing energy;
\item They are not highly correlated with each other.
\end{enumerate}
The following topological variables have been used to construct the likelihood function :
\begin{itemize}
\item Centrality = $\frac{H_T}{H}$ ; $H_{T}$ is the scalar sum of the $p_{T}$ of the four leading jets and $H$ is the scalar sum of the energy of the four leading jets, 
\item Aplanarity $A = \frac{3}{2}\times \lambda_{3}$ ; $\lambda_{3}$ is the smallest eigen value of the normalized momentum tensor
 $M_{ij} = \frac{\sum_{\circ}p_{i}^{\circ}p_{j}^{\circ}}{\sum_{\circ}\mid(\overrightarrow{p^{\circ}})\mid^{2}}$, where $\overrightarrow{p^{\circ}}$ is the momentum vector of reconstructed object $\circ$ ($e$, $\mu$ and four leading jets) and $i$ and $j$ are Cartesian coordinates. After proper diagonalization, $M_{ij} $ can have three eigenvalues, $\lambda_{1} \geq\lambda_{2}\geq\lambda_{3}$, with $\lambda_{1} +\lambda_{2} +\lambda_{3}=1$. Aplanarity is the measure of the flatness of the events. Spherical events should have large $A$ whereas planar events should have small $A$. $W+$jets events are more planar (small $A$) than $t\bar{t}$ events.

\item Sphericity $S= \frac{3}{2}\times(\lambda_{2}+\lambda_{3})$; where $\lambda_{2}$ and $\lambda_{3}$ are the smallest eigenvalues of the normalized momentum tensor ($\lambda_{3} < \lambda_{2}$) so that $0 \leq S\leq 1$. Sphericity is a measure of the summed $p_{T}^{2}$ with respect to the event axis. For isotropic events, $S \approx 1$ and for two jet events $S \approx 0$.
\item $\eta_{\ell}$ where $\ell = e, \mu$;
\item $\Delta\eta(j_2,j_3) = \eta(j_2) - \eta(j_3)$; 
\item $F (\theta_\ell, \theta_{j_1}) = 2\mid$ tan$^{-1}(e^{-\eta(\ell)})-$tan$^{-1}(e^{\eta(j_1)})\mid$; 


\item $F (\theta_\ell, \theta_{j_2}) = 2\mid$ tan$^{-1}(e^{-\eta(\ell)})-$tan$^{-1}(e^{\eta(j_2)})\mid$; 
\item $F (\theta_\ell, \theta_{j_3}) = 2\mid$ tan$^{-1}(e^{-\eta(\ell)})-$tan$^{-1}(e^{\eta(j_3)})\mid$; 


\end{itemize}
 The likelihood templates obtained using these topological variables are shown in Figure~1 and Figure~2.

\section{Expected statistical uncertainty}

In order to estimate the statistical uncertainty, we performed several ensemble tests where we mixed different proportions of $t\bar{t}$ and $W+$jets events to produce ``pseudo data''. We would like to see whether we can extract the exact fraction of $t\bar{t}$ events from the ``pseudo data'' using the default likelihood template. We have divided the $t\bar{t}$ samples into two halves. The first half of the $t\bar{t}$ samples was used for likelihood template construction. The second half of the $t\bar{t}$ sample was used for the ensemble tests. Due to the limited statistics of the $W+$jets samples, we used all available $W+$jets samples for both $W+$jets likelihood template construction and the ensemble test. We performed 1000 trials to obtain a distribution of the fitted signal fraction. Figure~3 and Figure~4 show the ensemble fit for the muon channel and electron channel respectively.

\begin{figure}
\begin{center}
\caption{\textbf{Likelihood templates for $\mu +$ jet  channel:  
$t\bar{t}$ template (solid red line), $W+$jets template (blue dashed line)}}
\includegraphics[width=2.6in]{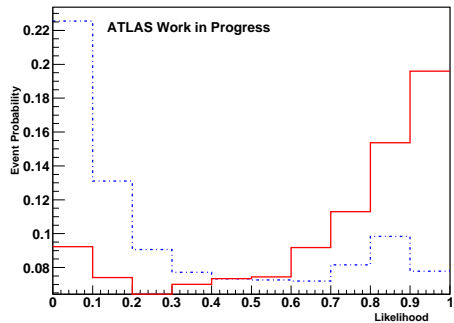}

\end{center}
\end{figure}


The initial $t\bar{t}$ fraction, average fitted fraction of $t\bar{t}$ events from the ensemble fit, and RMS of the distribution are presented in Table~III. There is good agreement between the input and fitted values of the $t\bar{t}$ fractions. The RMS of the distribution of the fitted fraction is considered to be the expected statistical error of the method. 

\begin{table}[htp]
\caption{\textbf{Initial $t\bar{t}$ fraction,  average $t\bar{t}$ fitted fraction and RMS for ensemble test}}
\begin{tabular}{| l |c |c | c|}
\hline
Decay Channel& Initial & Fitted & Statistical \\ 
      &fraction (\%) &fraction (\%) &error (\%)\\ \hline
\hline
$e$+jet& 69   & 68 & 6.0 \\ \hline
\hline
$\mu$+jet& 65  & 64 & 5.3 \\ \hline
\end{tabular}
\label{tab:fitted_fractions}
\end{table}

\section{Systematic uncertainties}

The systematic uncertainties on the $t\bar{t}$ production cross-section ($\Delta \sigma_{t\bar{t}}$) originate from uncertainties on the selection efficiency ($\Delta \epsilon_{sel}$) and uncertainties on $N_{t\bar{t}}$ ($\Delta N_{t\bar{t}}$, from the variations of the likelihood templates in the fit). Considering the fact that the systematic uncertainty from a given source can affect both the selection efficiency and the shape of the likelihood template at the same time, the systematic uncertainty on the $t\bar{t}$ production cross-section is determined by varying the source by one standard deviation up and down and propagating the variation into both fitted number of $t\bar{t}$ events and the signal efficiency [5]:

\begin{equation}
\sigma_{t\bar{t}} \pm \Delta \sigma_{t\bar{t}}= \frac{N_{t\bar{t}} \pm \Delta N_{t\bar{t}}}{Br \cdot L\cdot(\epsilon_{sel}\pm \Delta \epsilon_{sel})}
\end{equation}

Major sources of uncertainties due to jet energy scale, the Monte Carlo model used for the signal and background simulation, initial state radiation (ISR), final state radiation (FSR), and PDF are discussed below. 

\begin{figure}
\begin{center}
\caption{\textbf{Likelihood templates for $e+$jet channel :
$t\bar{t}$ template (solid red line), $W+$jets template (blue dashed line)}}
\includegraphics[width=2.6in]{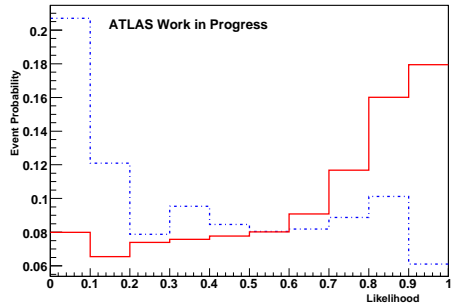}
\end{center}
\end{figure}

\subsection{Jet Energy Scale (JES)}
To obtain the uncertainty due to the jet energy scale, we varied all jet energies coherently by 10\% (pessimistic senario) and 5\% (optimistic senario). $\met$ has been corrected after varying the jet energy scale. We have seen that the likelihood template does not change significantly when we vary the jet energy scale. We calculated the modified selection efficiency for $t\bar{t}$ events after applying the jet energy scale. The change in selection efficiency due to variation of the jet energy scale propagates as a systematics uncertainity to the cross-section. 


To estimate the systematic uncertainty due to the likelihood shape, we derive new likelihood templates for  $t\bar{t}$ and $W+$jets after changing the jet energy. We make ``pseudo data'' using samples after varying the jet energies. Then we use the nominal likelihood template to find out the exact fraction of $t\bar{t}$ events from the ``pseudo data''. The difference between fitted fraction of $t\bar{t}$  events and the original fractions of $t\bar{t}$ events is considered to be the error due to the likelihood shape for JES. The systematic error due to the likelihood shapes for JES is negligible compared to the error due to selection efficiency. The contribution of JES systematics to the $\sigma_{t\bar{t}}$  has been summarized in Table ~IV.

\subsection{MC generators}
We calculated the uncertainty due to the MC generators in a similar manner as the JES systematic uncertainty. We used ACER MC to make a new ``pseudo data'' and applied ``nominal" templates to this data. The difference between the fitted fraction obtained on ACER  and MC\@@NLO events is considered to be the MC generator systematic uncertainty. We also calculated the modified selection efficiency for $t\bar{t}$ events and used it as the MC generator systematic uncertainty on the selection efficiency. The total systematic uncertainty due to MC generators is shown in Table~IV.    


\begin{figure}
\begin{center}
\caption{\textbf{Ensemble fit for $\mu +$ jet channel}}

\includegraphics[width=2.6in]{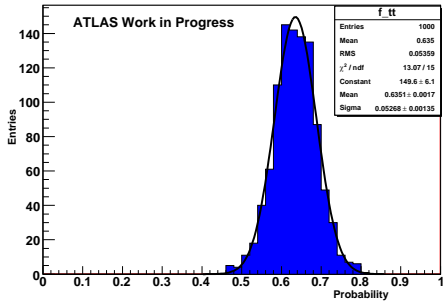}

\end{center}
\end{figure}

\subsection{ISR/FSR Modeling}


More initial state radiation (ISR) and final state radiation (FSR) increases the number of jets and affects the transverse momentum of particles. $t\bar{t}$ selection cuts include these quantities. Therefore, ISR and FSR will have an effect on the selection efficiency. In order to estimate the effect of the ISR/FSR systematics on the selection efficiency, we compared the results obtained with the $t\bar{t}$ AcerMC-generator [6] interfaced to PYTHIA [7] for showering and hadronisation. The Pythia parameters which control the ISR/FSR were then varied in order to generate the maximum difference for the reconstructed hadronic top masses and two reference samples created. The relative differences between the numbers of selected events obtained with these reference samples, relative to the default central sample, was taken as the uncertainty on the cross-section. In order to estimate the effect of ISR/FSR systematics on the likelihood shape, the nominal likelihood templates are obtained from half of the default central sample. Using the second half of the central sample and $W+$jets backgrounds, we produced ``pseudo data'', then we used the nominal likelihood template to find $t\bar{t}$  fractions from the ``pseudo data''.  A similar procedure was applied to two reference samples to find $t\bar{t}$  fractions from the ``pseudo data'' and the difference between the mean of the nominal sample to the mean of the reference samples was considered to be the uncertainty due to ISR/FSR. The systematic error due to ISR/FSR has been summarized in Table ~IV.


\subsection{Parton Distribution Function (PDF) uncertainties}
The uncertainties due to the PDF are examined by a re-weighting scheme that uses MC truth information about the hard partons. The basic procedure is to evaluate the probability of an event with a particular kinematic characteristic to be produced [4]. The variation of the selection efficiency on the signal sample was taken as a measure of the cross-section measurement variation [8]. The errors coming from the PDF uncertainties are shown in Table~IV. We did not evaluate the PDF uncertainty due to the change of the likelihood shape since this error is significantly smaller compared to others, like JES, ISR/FSR and Monte Carlo generators.

\begin{figure}
\begin{center}
\caption{\textbf{Ensemble fit for $e +$ jet channel}}

\includegraphics[width=2.6in]{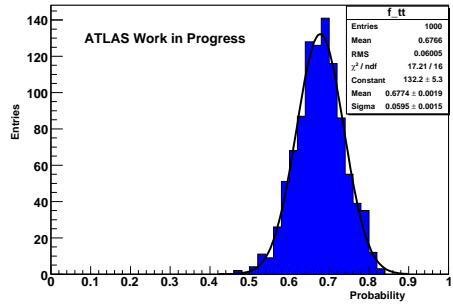}


\end{center}
\end{figure}

\subsection{Trigger efficiency}
The lepton trigger efficiency is measured from the simulated MC data using Z events. We expect the uncertainty to be of the order of 1\% for electrons and muons [4].

\subsection{Single top}
The systematic uncertainty due to the single top production cross-section is shown in Table IV. Its contribution to $t\bar{t}$ and background templates is calculated by varying the theoretical cross-section by $\pm 12\%$. 

\subsection{Luminosity}
An uncertainty of 20$\%$ on the integrated luminosity is assumed for the early LHC data-taking period. The luminosity error is directly translated into the cross-section error and results in $\pm$22$\%$ for the $t\bar{t}$ cross-section measurement [4].




Tables~IV summarizes the contributions from each source of the considered systematic uncertainties on the selection efficiency as well as due to changes in the likelihood shapes. One can see that the largest contribution comes from the JES. The total systematic uncertainty is calculated as a quadrature sum of uncertainties due to various sources.


\begin{table}[htp]
\caption{\textbf{Summary of systematic uncertainties}}
\begin{tabular}{|c |c |c |}
\hline
Source of systematics & $\mu +$ jets channel & $e +$ jets channel  \\ 
& ($\%$)  & ($\%$) \\\hline
Trigger &      $\pm$1.0  & $\pm$1.0 \\ \hline
Lepton ID &    $\pm$1.0 & $\pm$1.0 \\ \hline
JES(5\% up) &   +9.7     & +10 \\ \hline
JES(5\% down) & -11.1   &  -10 \\ \hline
JES(10\% up) & +19.4    &  +20 \\ \hline
JES(10\% down) &-20.8       & -21.6 \\ \hline
MC Generator & $\pm$5.0  & $\pm$3.5 \\ \hline
ISR & +7.7  & +10.2 \\ \hline
FSR & -7.9  & -9.1 \\ \hline
PDF & $\pm$1.2  & $\pm$1.6 \\ \hline
Single top & $\pm$0.5  & $\pm$0.7 \\ \hline
Luminosity & $\pm$22  & $\pm$22 \\ \hline

\end{tabular}
\label{tab:syst_err_1}
\end{table}




\section{Conclusions}

The $t\bar{t}$ cross-section measurement using the likelihood method has been discussed. During early LHC operation the ATLAS detector will not be well understood. No b-tagging is used in this analysis. We have chosen likelihood variables that do not depend on jet energy scale and missing energy so that we can use them during early LHC operation and we can minimize systematic uncertainties. As a result, the dominant systematic error (jet energy scale) is reduced by a factor of two compared to a cut-based analysis [4]. The $t\bar{t}$ cross-section measurement at 50 pb$^{-1}$ using the likelihood method can be measured with the following accuracy:
\begin{itemize}
\item \textbf{$\mu+$jets channel (for $\pm$ 5$\%$ JES)}:
\begin{equation}
\frac{\Delta \sigma_{t\bar{t}}}{\sigma_{t\bar{t}}} = \pm 5.3\% (stat.)~^{+15.6}_{-16.5} \% (syst.) \pm22\% (lumi).
\end{equation} 

\item \textbf{$\mu+$jets channel (for $\pm$ 10$\%$ JES)}:
\begin{equation}
\frac{\Delta \sigma_{t\bar{t}}}{\sigma_{t\bar{t}}} = \pm 5.3\% (stat.)~^{+22.9}_{-24.1} \% (syst.) \pm22\% (lumi).
\end{equation} 


\item \textbf{$e+$jets channel (for $\pm$ 5$\%$ JES)}:
\begin{equation}
\frac{\Delta \sigma_{t\bar{t}}}{\sigma_{t\bar{t}}} = \pm 6.0\% (stat.)~^{+17.4}_{-17.4} \% (syst.) \pm22\% (lumi).
\end{equation}

\item \textbf{$e+$jets channel (for $\pm$ 10$\%$ JES)}:
\begin{equation}
\frac{\Delta \sigma_{t\bar{t}}}{\sigma_{t\bar{t}}} = \pm 6.0\% (stat.)~^{+24.5}_{-25.8} \% (syst.) \pm22\% (lumi).
\end{equation}

\end{itemize}


\begin{acknowledgments}
The author would like to thank the organizers of DPF 2009 conference for creating a fruitful collaborating environment. The author would like to thank P. Skubic, B. Abbott, F. Rizatdinova, A. Khanov, T. Golling, F. Spano, M. Saleem, B. Abi for their help and suggestions at the various stages of this analysis.
\end{acknowledgments}

\end{document}